\documentclass[conference]{IEEEtran}
\IEEEoverridecommandlockouts
\usepackage{cite}
\usepackage{amsmath,amssymb,amsfonts}
\usepackage{algorithmic}
\usepackage{listings}
\usepackage{graphicx}
\usepackage{booktabs}
\usepackage{tabularx}
\usepackage{multirow}
\usepackage{enumitem}
\usepackage{textcomp}
\usepackage[absolute]{textpos}
\usepackage[dvipsnames]{xcolor}
\usepackage[a4paper, total={184mm,239mm}]{geometry}
\def\BibTeX{{\rm B\kern-.05em{\sc i\kern-.025em b}\kern-.08em
    T\kern-.1667em\lower.7ex\hbox{E}\kern-.125emX}}
\usepackage{glossaries}
\usepackage[hidelinks]{hyperref}
\usepackage[capitalise,nameinlink]{cleveref}
\usepackage{microtype}
\usepackage[free-standing-units,per-mode=repeated-symbol,mode=text,detect-weight=true,detect-family=true]{siunitx}
\usepackage[shortcuts]{extdash}
\DeclareSIUnit\core{core}
\DeclareSIUnit\tile{tile}
\DeclareSIUnit\request{request}
\DeclareSIUnit\cycle{cycle}
\DeclareSIUnit\mac{MACs}
\DeclareSIUnit\erlang{E}
\DeclareSIUnit\flop{FLOP}
\DeclareSIUnit\flops{FLOPS}
\DeclareSIUnit\gate{GE}
\DeclareSIUnit\ge{GE}
\DeclareSIUnit\op{OP}
\DeclareSIUnit\ops{OPS}
\DeclareSIUnit\bps{bps}
\DeclareSIUnit\Bps{Bps}
\DeclareSIUnit\ipc{IPC}
\DeclareSIUnit[number-unit-product = ]\percent{\%}

\newacronym{abi}{ABI}{application binary interface}
\newacronym{ace}{ACE}{AXI Coherent Extensions}
\newacronym{alu}{ALU}{arithmetic logic unit}
\newacronym{amba}{AMBA}{Advanced Microcontroller Bus Architecture}
\newacronym{amo}{AMO}{atomic memory operation}
\newacronym{aot}{AOT}{ahead-of-time}
\newacronym{apb}{APB}{Advanced Peripheral Bus}
\newacronym{api}{API}{application programming interface}
\newacronym{asic}{ASIC}{application-specific integrated circuit}
\newacronym{axi}{AXI}{Advanced eXtensible Interface}
\newacronym{blas}{BLAS}{Basic Linear Algebra Subprograms}
\newacronym{cas}{CAS}{compare-and-swap}
\newacronym{cgra}{CGRA}{coarse-grained reconfigurable architectures}
\newacronym{cmos}{CMOS}{complementary metal-oxide-semiconductor}
\newacronym{cnn}{CNN}{convolutional neural network}
\newacronym{cpu}{CPU}{central processing unit}
\newacronym{csr}{CSR}{control and status register}
\newacronym{dbt}{DBT}{dynamic binary translation}
\newacronym{dct}{DCT}{discrete cosine transform}
\newacronym{dlp}{DLP}{data level parallelism}
\newacronym{dma}{DMA}{direct memory access}
\newacronym{dram}{DRAM}{dynamic random-access memory}
\newacronym{dsl}{DSL}{domain-specific language}
\newacronym{dsp}{DSP}{digital signal processing}
\newacronym{elf}{ELF}{Executable and Linkable Format}
\newacronym{fdsoi}{FD-SOI}{fully depleted silicon-on-insulator}
\newacronym{fifo}{FIFO}{first in, first out}
\newacronym{fpga}{FPGA}{field-programmable gate array}
\newacronym{fpu}{FPU}{floating-point unit}
\newacronym{fsm}{FSM}{finite-state machine}
\newacronym{gpgpu}{GPGPU}{general-purpose computing on a \gls{gpu}}
\newacronym{gpu}{GPU}{graphics processing unit}
\newacronym{hart}{hart}{hardware thread}
\newacronym{hbm}{HBM}{High Bandwidth Memory}
\newacronym{hdl}{HDL}{hardware description language}
\newacronym{hero}{HERO}{Heterogeneous Embedded Research Platform}
\newacronym{hpc}{HPC}{high-performance computing}
\newacronym{ilp}{ILP}{instruction level parallelism}
\newacronym{ipc}{IPC}{instructions per cycle}
\newacronym{iot}{IoT}{Internet of Things}
\newacronym{ipu}{IPU}{image processing unit}
\newacronym{ir}{IR}{intermediate representation}
\newacronym{isa}{ISA}{instruction set architecture}
\newacronym{issr}{ISSR}{indirection stream semantic register}
\newacronym{jit}{JIT}{just-in-time}
\newacronym{llc}{LLC}{last-level cache}
\newacronym{lrsc}{LRSC}{load-reserved/store-conditional}
\newacronym{lr}{LR}{load-reserved}
\newacronym{lsu}{LSU}{load-store unit}
\newacronym{mac}{MAC}{multiply–accumulate}
\newacronym{mimd}{MIMD}{multiple instruction, multiple data}
\newacronym{mmu}{MMU}{memory management unit}
\newacronym{mwait}{MWait}{Memory Wait}
\newacronym[longplural={networks-on-chip}]{noc}{NoC}{network-on-chip}
\newacronym{numa}{NUMA}{non-uniform memory access}
\newacronym{pc}{PC}{program counter}
\newacronym{pe}{PE}{processing element}
\newacronym{pl}{PL}{programmable logic}
\newacronym{pmca}{PMCA}{programmable manycore accelerator}
\newacronym{pulp}{PULP}{Parallel Ultra Low Power}
\newacronym{qlr}{QLR}{queue-linked register}
\newacronym{raw}{RAW}{read-after-write}
\newacronym{rom}{ROM}{read-only memory}
\newacronym{rmw}{RMW}{read–modify–write}
\newacronym{rob}{ROB}{reorder buffer}
\newacronym{ro}{RO}{read-only}
\newacronym{rtl}{RTL}{register-transfer level}
\newacronym{sbt}{SBT}{static binary translation}
\newacronym{scm}{SCM}{standard cell memory}
\newacronym{sc}{SC}{store-conditional}
\newacronym{sdf}{SDF}{Standard Delay Format}
\newacronym{simd}{SIMD}{single instruction, multiple data}
\newacronym{simt}{SIMT}{single instruction, multiple thread}
\newacronym{sm}{SM}{streaming multiprocessor}
\newacronym{soc}{SoC}{system-on-chip}
\newacronym[longplural={scratchpad memories}]{spm}{SPM}{scratchpad memory}
\newacronym{sram}{SRAM}{static random-access memory}
\newacronym{ssa}{SSA}{static single assignment}
\newacronym{ssr}{SSR}{stream semantic register}
\newacronym{tcdm}{TCDM}{tightly-coupled data memory}
\newacronym{tlp}{TLP}{thread-level parallelism}
\newacronym{vpu}{VPU}{vector processing unit}
\newacronym{vliw}{VLIW}{very long instruction word}
\newacronym{vnb}{VNB}{von Neumann bottleneck}
\newacronym{war}{WAR}{write-after-read}
\newacronym{waw}{WAW}{write-after-write}
\newacronym{cs}{CS}{critical section}
\newacronym{mpmc}{MPMC}{multi producer, multi consumer}
\newacronym{mcs}{MCS}{Mellor-Crummey, Scott}
\newacronym{tas}{TAS}{test-and-set}
\newacronym{rsf}{RSF}{Request-Store-Forward}
\newacronym{qnode}{Qnode}{queue node}

\PackageWarning{Citation missing:}{#1!}
\PackageWarning{TODO:}{#1!}

\newcommand\etal{et\penalty50\ al.}
\newcommand\riscv{RISC\=/V}
\newcommand\mempool{Mem\-Pool}
\newcommand\lrwait{LRwait}
\newcommand\scwait{SCwait}
\newcommand\lrscwait[1]{LRSCwait\ensuremath{_{\text{#1}}}}

\newcommand{\lrsc}{\gls{lrsc}}
\newcommand{\mwait}{Mwait}
\newcommand{\qnode}{\gls{qnode}}
\newcommand{\colibri}{Colibri}

\newcommand\circledigit[1]{{(#1)}}

\begin{document}
\bstctlcite{IEEEexample:BSTcontrol}

\title{\lrscwait{}: Enabling Scalable and Efficient Synchronization in Manycore Systems through Polling-Free and Retry-Free Operation}

\author{%
Samuel Riedel\textsuperscript{\ensuremath{*}}\quad
Marc Gantenbein\textsuperscript{\ensuremath{*}}\quad
Alessandro Ottaviano\textsuperscript{\ensuremath{*}}\quad
Torsten Hoefler\textsuperscript{\ensuremath{\dag}}\quad
Luca Benini\textsuperscript{\ensuremath{*\ddagger}}
\\
{\small
 \textsuperscript{\ensuremath{*}}IIS, ETH Z\"{u}rich\quad%
 \textsuperscript{\ensuremath{\dag}}SPCL, ETH Z\"{u}rich\quad%
 \textsuperscript{\ensuremath{\ddag}}DEI, University of Bologna%
}
\\
{\small\itshape%
 \textsuperscript{\ensuremath{*}}\{sriedel,ganmarc,aottaviano,lbenini\}@ethz.ch\quad%
 \textsuperscript{\ensuremath{\dag}}htor@inf.ethz.ch%
}
}

\begin{textblock}{0.8}[0.5,0.5](0.5,0.93) 
  \begin{center} 
    \tiny
    \copyright~2024 IEEE. Personal use of this material is permitted. Permission from IEEE must be obtained for all other uses, in any current or future media, including reprinting/republishing this material \\ for advertising or promotional purposes, creating new collective works, for resale or redistribution to servers or lists, or reuse of any copyrighted component of this work in other works.  
  \end{center}
\end{textblock}

\maketitle

\begin{abstract}
Extensive polling in shared-memory manycore systems can lead to contention, decreased throughput, and poor energy efficiency. Both lock implementations and the general-purpose atomic operation, \gls{lrsc}, cause polling due to serialization and retries. To alleviate this overhead, we propose \lrwait{} and \scwait{}, a synchronization pair that eliminates polling by allowing contending cores to sleep while waiting for previous cores to finish their atomic access. As a scalable implementation of \lrwait{}, we present \colibri{}, a distributed and scalable approach to managing \lrwait{} reservations. Through extensive benchmarking on an open-source \riscv{} platform with 256 cores, we demonstrate that \colibri{} outperforms current synchronization approaches for various concurrent algorithms with high and low contention regarding throughput, fairness, and energy efficiency. With an area overhead of only \SI{6}{\percent}, \colibri{} outperforms \gls{lrsc}-based implementations by a factor of 6.5$\times$ in terms of throughput and 7.1$\times$ in terms of energy efficiency.
\end{abstract}


\glsresetall

\section{Introduction}

Manycore systems are becoming increasingly popular due to the growing demand for computing power. However, the parallel execution of tasks introduces synchronization and atomicity issues that can lead to race conditions and unpredictable results. To ensure exclusive access to \glspl{cs}, atomic operations and locks can be used. However, locks also block cores that try to acquire them when they are not free, leading to busy waiting and polling. Polling, or constantly checking a shared resource for changes, can become an issue in concurrent algorithms. It leads to high core utilization and reduces overall system performance and energy efficiency as the cores compete for shared resources~\cite{Anderson1990}. In the worst case, it can lead to livelocks or starvation, where cores are blocked from making progress because others continuously block them.

Non-blocking algorithms avoid locks by updating atomic variables directly with atomic \gls{rmw} operations. Specific arithmetic operations, like \emph{add, and, or}, are often supported through specialized instructions. However, most concurrent algorithms require more complex modifications of atomic variables, such as conditional updates. For generic \gls{rmw} operations, the \gls{cas} operations or \gls{lrsc} pair are typical primitives designed to ensure that the operation is \emph{atomic}, i.e., without interference from other cores~\cite{Herlihy2020}. For example, \riscv{}'s \gls{lr} instruction loads a value from memory and places a reservation. The core can perform operations with the loaded value and store the result back conditionally with a \gls{sc}. The latter instruction will only succeed if the reservation is still valid, meaning the memory location was not modified in the meantime. If the \gls{sc} succeeds, the \gls{rmw} sequence appears atomically. However, cores that fail an \gls{sc} must retry the \gls{lrsc} sequence pair until it succeeds. Variables outside \glspl{cs} can also cause polling, where cores wait for changes in shared variables, leading to inefficiencies in core communication, like producer/consumer interactions.

To eliminate retries and polling, we propose a novel, general-purpose atomic \gls{rmw} instruction pair called \lrwait{} and \scwait{}. They extend the standard \riscv{} \gls{lrsc} pair by moving the linearization point, the point where the atomic operations of different cores get ordered, from the \gls{sc} to the \lrwait{}. The \lrwait{} and \scwait{} are used in the same way as the \gls{lrsc} pair. However, instead of returning the memory value immediately, the \lrwait{} instruction only responds to one core at a time to set it up for a successful \scwait{}. This prevents failing \glspl{sc} and retry loops. Furthermore, \lrscwait{} allows implementing polling-free locks. To eliminate polling even for non-atomic variables, we propose the \mwait{} instruction, which enables cores to sleep until a specific memory address changes its value.

While cache-based systems often rely on the coherency protocol to implement such behavior, manycore accelerators scaling to hundreds of cores often rely on software-managed, multi-banked \glspl{spm}. Examples include commercial chips like GAP9~\cite{GreenWavesTechnologies2021} and RC64~\cite{Ginosar2016}, as well as large-scale research prototypes like \mempool{}~\cite{Riedel2023}. While \lrscwait{} can be applied to cache and cache-less systems, in this work, we focus on cache-less, \gls{spm}-based manycore systems since they pose the design challenge of the memory controllers having to keep track of outstanding \lrwait{} instructions to send their responses at the right time. However, duplicating large hardware queues for each bank is costly and scales poorly.

As a scalable implementation of the proposed instructions, we present \emph{\colibri{}}. Its concept is similar to linked-list-based software queues. It does not allocate a full array of entries for each queue but just a head and tail pointer per queue as illustrated in \cref{fig:architectures}. Each core is equipped with a queue node that can be linked to any queue. For \colibri{}, this means that instead of equipping each memory controller with a hardware queue that can hold an entry for each core, each memory controller is extended with a parameterizable number of head and tail registers to form linked lists. Each core is then equipped with one hardware queue node, and when issuing an \lrwait{}, the core inserts itself in the corresponding queue. We implemented \colibri{} on the open-source, manycore \mempool{} system, consisting of 256 cores sharing \SI{1}{\mebi\byte} of L1 memory~\cite{Riedel2023}. \colibri{} provides a scalable solution that can be easily integrated into existing \riscv{} systems. The \lrscwait{} solution can be used as a drop-in replacement for \gls{lrsc} or as a powerful extension, making it a desirable option for high-performance computing systems. We evaluate the performance of \colibri{} against various hardware and software approaches. The results indicate that \colibri{} outperforms other approaches in all experiments, with a throughput increase of up to 6.5 times in high-contention situations and a \SI{13}{\percent} increase in low-contention scenarios. Additionally, \colibri{} reduces polling, allowing other applications to be unaffected by concurrent atomic accesses. Our key contributions are the following:

\begin{itemize}
    \item The \lrwait{} extension consisting of three novel instructions (\lrwait{}, \scwait{}, and \mwait{}), which enable atomic access and monitoring memory locations with a minimal amount of polling (\cref{sec:lrwait}).
    \item A scalable implementation for \lrwait{} named \colibri{} leveraging a distributed reservation queue (\cref{sec:colibri}).
    \item An implementation and evaluation of \colibri{} on the \mempool{} platform that outperforms other approaches in throughput, fairness, polling, and energy per atomic access. \colibri{} scales linearly on the \mempool{} platform by introducing an area overhead of just \SI{6}{\percent} while being 8.8x more energy efficient than locks (\cref{sec:results}).
\end{itemize}


\section{Related Work}

A common approach to mitigate polling is using a backoff after a failed atomic access~\cite{Herlihy2020}. Existing backoff schemes, such as exponential backoff, where each failed attempt increases the backoff time, can reduce the overhead on shared resources but still make the cores busy-waiting and performing sub-optimally.

The \gls{mcs} lock~\cite{Mellor-Crummey1991} relies on a software queue for contending cores to enqueue in and spin on their respective node in the queue. This guarantees that each core spins on a unique location to mitigate contention on the lock variable itself. This approach works well in cache-based systems since each core can spin on its own L0 cache. However, in this work, we focus on systems with software-managed memories.

While software approaches to locks are general and platform agnostic, their performance can not keep up with hardware locks. A study of two software locks and four hardware locks shows that hardware locks consistently outperform the software approaches by \SI{25}{\percent}-\SI{94}{\percent}. However, hardware locks such as Hardlocks~\cite{Strom2019} do not scale well, as the locks are managed by a centralized locking unit accessible to all cores. Accessing this unit quickly becomes the bottleneck in large systems. Furthermore, the number of locks is fixed at implementation time. Similarly, Glaser \etal{} present a synchronization unit where each core has a private direct connection to each hardware lock~\cite{Glaser2021}. While this solves the contention issue, it prevents scaling beyond a few tens of cores. GLock suffers from a similar scalability issue~\cite{Abellan2013}. It is based on a dedicated on-chip network consisting of lock managers and local controllers that synchronize to acquire a lock. Monchiero \etal{} propose a synchronization-operation buffer implemented as a hardware queue in the memory controller to resolve the lock accesses~\cite{Monchiero2006}. However, this approach only implements locks and has a hardware cost that is proportional to the number of cores. Furthermore, each memory controller would require such a buffer to manage locks.

While locks are a common solution for protecting critical sections, their blocking nature often limits performance. Lock-free algorithms, on the other hand, allow for much more concurrency. They often rely on instructions like \gls{cas} or the \gls{lrsc} pair. This section focuses on the latter, specifically, \riscv{}'s implementation. For example, the ATUN is a unit that can be placed in an \gls{axi} bus to support \gls{lrsc} instructions to the downstream memory~\cite{Kurth2020}. The table allows a reservation for every core, thus implementing a non-blocking version of \gls{lrsc}. Furthermore, each bank would require its own ATUN adapter in a multi-banked system, introducing significant hardware overhead in large manycore systems. The Rocket chip features a similar implementation~\cite{Asanovic2016}. However, the number of reservations is limited.

\mempool{} implements a lightweight version of \gls{lrsc} by only providing a single reservation slot per memory bank~\cite{Riedel2023}. However, this sacrifices the non-blocking property of the \gls{lrsc} pair. The GRVI multiprocessor, on the other hand, modifies the granularity at which \glspl{lrsc} operate by locking the complete memory bank~\cite{Gray2016}. This reduces the hardware overhead to one bit per core per bank, albeit the approach is still affected by retries due to spuriously failing \gls{sc} operations.

All those solutions implement the standard \riscv{} \gls{lrsc} instruction, leveraging the freedom of the official specification to achieve different trade-offs. However, none of them solve the polling and retry issue of failing \gls{sc} operations. On the contrary, they sometimes worsen it. The \gls{rsf} synchronization model proposed by Liu \etal{} is similar to \lrwait{}~\cite{Liu2007}. Synchronization requests are stored in a hardware-managed memory and handled in order by a synchronization controller. However, this approach leads to a high memory footprint, and the hardware needs to be replicated for each memory bank. Furthermore, it is infeasible for software-managed memories as the synchronization controller will interfere with the allocated data when adding the queue to the memory.

Our \lrwait{} approach and the efficient implementation through \colibri{} scale well to hundreds of cores and banks while completely eliminating polling without sacrificing granularity.


\section{\lrwait{} and \scwait{}}%
\label{sec:lrwait}

\riscv{} defines the \acrfull{lrsc} instructions to implement generic, atomic \gls{rmw} operations. The \gls{lr} instruction reads a value from memory and places a reservation, which remains valid until the specified memory address is changed. The core can then modify the value and write the result back with an \gls{sc} instruction. The latter will succeed only if the reservation is still valid. If the \gls{sc} fails, the \gls{lrsc} sequence has to be retried. The linearization point between contending cores is thus at the \gls{sc}.

\lrwait{} eliminates the wasteful retry loop by moving the linearization point to the \lrwait{} instruction, i.e., atomic accesses of competing cores are ordered at the \lrwait{} instruction. Instead of immediately returning the value, the memory controller withholds the response such that only one core gets a response at a time, guaranteeing it to be the only core issuing an \scwait{} to the same address. The \lrscwait{} and \gls{lrsc} instructions share similar semantics. The \scwait{} stores a value conditionally and returns a success or failure code analogous to the \gls{sc}. Likewise, the \lrwait{} matches the \gls{lr} instruction, but its response is delayed. The sequence of an atomic \gls{rmw} operation with \lrscwait{} is the following:

\begin{enumerate}
    \item The core issues the \lrwait{} and waits for the response.
    \item The memory buffers the request until it is the next outstanding \lrscwait{} pair to that address.
    \item Once the \lrwait{} is the next in line, the memory serves the request with the current memory value and monitors it. A store to the same address clears the reservation.
    \item The core modifies the value and writes it with an \scwait{}.
    \item The memory accepts the value if a valid reservation still exists and issues the response.
\end{enumerate}

While the memory guarantees that only one core proceeds with an \lrscwait{} pair, it cannot eliminate the possibility of another core overwriting the atomic variable, leading to a failing \scwait{}. 
One constraint of the \lrscwait{} instruction pair is that every \lrwait{} must eventually be followed by an \scwait{}. While \riscv{} does not have this constraint for \gls{lrsc}, our extension requires the matching \scwait{} to yield the queue of outstanding \lrwait{} instructions and allow progress on the atomic variable. Albeit \lrscwait{} can be used as a drop-in replacement for \gls{lrsc}, it removes the lock-free progress guarantee that the \gls{lrsc} instructions have. Since only one core can issue an \scwait{}, a malicious core could block the resource indefinitely and obstruct progress. However, \lrscwait{} still gives strong progress guarantees under the following constraints:

\paragraph{Mutual exclusion}
Just as the \lrsc{} pair, the \scwait{} only succeeds if a valid reservation is present, meaning there was no write between the \lrwait{} and the \scwait{}, which guarantees mutual exclusion and, therefore, atomicity. 

\paragraph{Deadlock freedom}
To prevent circular dependencies between cores, every core must have at most one outstanding \lrwait{} operation. \riscv{} does not impose this requirement on \lrsc{}. However, only the innermost \lrsc{} pair is guaranteed to progress. Therefore, this requirement for deadlock freedom is a requirement for livelock freedom already. Furthermore, each core's \lrwait{} must eventually be followed by an \scwait{} to close the \gls{cs}. We impose the same constraints as the \riscv{} standard to allow only a finite and limited set of instructions between \lrwait{} and \scwait{}.

\paragraph{Starvation freedom}
Starvation freedom guarantees that all cores eventually make progress. \Gls{lrsc} only guarantees that one core makes progress because an unlucky core could always lose the \gls{sc} to a faster core. In our work, this scenario is prevented by handling the \lrscwait{} pairs in order, thus enabling all cores to eventually execute the \lrscwait{} pair and, therefore, guaranteeing starvation freedom.

Overall, while the blocking nature of the \lrscwait{} makes a core's progress depend on other cores correctly executing and leaving the \lrscwait{} blocks, these constraints can easily be adhered to in bare-metal systems, which are fully under the programmer's control. \lrscwait{} can provide very strong progress guarantees, enabling each core to progress. However, hardware failure or software bugs can become blocking.

\subsection{Ideal Hardware Implementation}\label{subsec:ideal-hw}

\begin{figure}[t]
    \centering
    \includegraphics[]{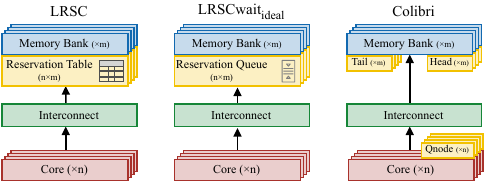}
    \caption{Difference between \gls{lrsc} architecture with a reservation table, \lrscwait{} with a reservation queue, and \colibri{} with a linked-list-like structure.}%
    \label{fig:architectures}
\end{figure}

A straightforward hardware implementation of \lrscwait{} requires tracking all outstanding reservations in order to ensure fairness and starvation freedom. As shown in \cref{fig:architectures}, this can be achieved by an \lrscwait{} adapter placed in front of each memory bank, consisting of (i) a queue-like data structure of capacity $n$, where $n$ is the number of cores in the system, and (ii) some additional logic to monitor memory accesses and invalidate reservations when the target address is overwritten. The overhead of this implementation in a system with $m$ memory banks is $O(n\log_2(n)m)$, where $\log_2(n)$ represents identifier size per core. Assuming that $m$ scales linearly with the number of cores, this implementation's overhead scales quadratically with the system size: $O(n^2)$, a non-negligible hardware complexity.

\subsection{Optimized Hardware Implementation}

To reduce the hardware complexity, we can decrease the queue's capacity by assuming that only a subset of cores can access a specific address simultaneously. Our implementation supports a parametrizable number of reservation slots $q$. The case with $q=n$ falls back to the ideal \lrscwait{} pair described in \cref{subsec:ideal-hw}. We call this implementation \emph{\lrscwait{ideal}}. If $q<n$, we trade hardware overhead with performance. In these implementations, \emph{\lrscwait{q}}, cores executing an \lrwait{} to a full queue will fail immediately.

\subsection{\mwait{}}

To allow efficient monitoring and notification of a memory location from a core in the system, we introduce \emph{\mwait{}}. \mwait{} is derived from \lrwait{}, but without a matching \scwait{}. Instead, the reservation placed by \mwait{} is used to identify the core that needs to be notified of a change. 
For instance, a core may monitor a queue and be woken up when an element is pushed onto the queue.
Our experiments show that \mwait{} provides a simple and efficient mechanism for monitoring memory locations, allowing cores to be woken up only when necessary. To handle the possibility that the change we wish to observe has already occurred, we provide \mwait{} with an expected value. If the memory location already differs from the expected value when \mwait{} is served, the core is immediately notified.

\section{\colibri{}}%
\label{sec:colibri}

\colibri{} implements a distributed queue, similar to a linked list, shown in \cref{fig:architectures}. It alleviates the huge hardware overhead of the hardware queues at each memory controller, replacing it with a dedicated head and tail node per queue and a simple controller. On top of that, each core requires its own hardware node, called \emph{\qnode{}}, to enqueue itself. Since each core can only be in one queue, one \gls{qnode} per core is enough. Therefore, \colibri{} only requires $O(n+2m)$ nodes and scales linearly with the system size.

Since the queue is distributed across \glspl{qnode} and the head/tail nodes next to the memory banks, updating the queue becomes more complex. In comparison to the ideal \lrwait{}, an enqueue operation from an \lrwait{}, or a dequeue operation by an \scwait{}, does not happen in one place and a single cycle.

\begin{figure}[b]
    \centering
    \includegraphics[]{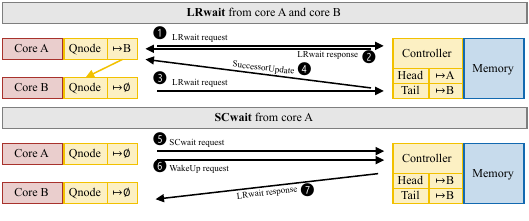}
    \caption{\lrwait{} and \scwait{} sequence in \colibri{} with two cores and one queue.}%
    \label{fig:colibri}
\end{figure}

We present a simple example of the construction and deconstruction of the queue in \cref{fig:colibri} with a single memory and two cores contending for the same address. Both cores have their own \glspl{qnode}, and the memory has a head and tail node. We call the cores \emph{A} and \emph{B} for simplicity.

\paragraph{\lrwait{}}
Core A issues an \lrwait{} request to the memory \circledigit{1}. Since the queue is initially empty, the head and tail nodes are set to A, and a reservation to the specified location is set up. The memory then sends the value A \circledigit{2}. During or after the described events, B's \lrwait{} request arrives at the memory \circledigit{3}. When the B's \lrwait{} request arrives at the memory, the controller appends B at the tail of the queue and then adds it as the successor to A. This is done by sending a so-called \emph{SuccessorUpdate} to A \circledigit{4}. This SuccessorUpdate writes to A's \gls{qnode} to make it point to B. In this final state shown in the top half of \cref{fig:colibri}, A and B form a queue with A at the head of the queue. At this point, A can issue an \scwait{} while B is sleeping, waiting for a response.

\paragraph{\scwait{}}
Core A finishes its \lrscwait{} pair by issuing an \scwait{} with the modified value \circledigit{5}. Immediately after an \scwait{} passes the \qnode{}, it sends a \emph{WakeUpRequest} to the memory containing its successor, i.e., B \circledigit{6}. On arrival of the \scwait{} request at the memory, the head node and reservation are checked. If everything is valid, the head node is temporarily invalidated to prevent a future \scwait{} from the same core from succeeding without reservation, and the \scwait{} is written to memory. The WakeUpRequest sets the head node to the successor node and triggers an \lrwait{} response with the latest memory value written by A, i.e., for B \circledigit{7}. Core B is now free to issue an \scwait{}. Finally, the head and tail nodes point to B since B is the only core in the queue.

This sequence can be generalized to more cores. \Glspl{qnode} accept SuccessorUpdates even when the core is asleep, allowing the queue to be enlarged independent of the cores' state.

\subsection{Correctness of \colibri{}}

\subsubsection{\lrwait{}}

When an \lrwait{} enqueues a node, it must update the tail to point to the newly enqueued node and append it to the previous tail node if it existed. If not, the enqueue operation inherently becomes atomic. Otherwise, to update the predecessor, the memory controller sends a SuccessorUpdate to the previous tail and overwrites the tail node atomically. Since we can only have one \lrwait{} per core and SuccessorUpdates are only sent when overwriting a tail node, only a single SuccessorUpdate will ever be in flight to a \gls{qnode}, guaranteeing no lost links in the queue. If the SuccessorUpdate arrives after the core issued an \scwait{}, it will immediately bounce back as a WakeUpRequest. If the next \lrwait{} arrives while the SuccessorUpdate is still in flight, the tail will be updated again, and the SuccessorUpdate will be sent to the next core. While a glance at the \glspl{qnode} might reveal broken links momentarily, the links only have to be made when a core issues its \scwait{}, which requires an \lrwait{} response from the memory controller since memory transactions are ordered, this will always happen after the SuccessorUpdate.

\subsubsection{\scwait{}}

If a core issuing an \scwait{} is the only one in the queue, i.e., the head and tail are equal, dequeuing itself by clearing the head and tail is trivial. Otherwise, the \scwait{} will invalidate the head node while leaving the value unchanged. A core would need to overwrite the head node to reach an inconsistent queue from this stage. This is only allowed for an \lrwait{} reaching an empty queue or a WakeUpRequest arriving at the memory after invalidating the head node by an \scwait{}. A WakeUpRequest can only be triggered by an \scwait{} passing the \gls{qnode}, which can only be sent by a core at the head of the queue since the other cores are still waiting for their \lrwait{} response. Thus, the WakeUpRequest arriving at the memory node guarantees that the queue is in a consistent state again.

\subsection{Extending \colibri{} with \mwait{}}

A core can issue an \mwait{} request to enqueue into \colibri{}'s queue to monitor a memory location. The memory controller then waits for a write to the monitored location, just like for \lrwait{}'s reservation. After a write, the memory controller triggers a response to the \mwait{} instruction. For \mwait{}, the head node is sleeping as well in contrast to \lrscwait{} where the head is free to issue an \scwait{}. The \mwait{} response makes the \gls{qnode} dispatch the WakeUpReq for its successor, which then bounces to the memory controller, where the next \mwait{} response is released. In contrast to \lrscwait{}, the whole reservation queue is woken up without any interference from the cores.

\section{Results}%
\label{sec:results}

We implement and evaluate various \lrscwait{} variations and \colibri{} in \mempool{}, an open-source, 256-core \riscv{} system with 1024 \gls{spm} banks~\cite{Riedel2023}. All our results are taken from cycle-accurate \gls{rtl} simulation. Physical implementation results come from implementing \mempool{} in GlobalFoundries' 22FDX \gls{fdsoi} technology. Power consumption is evaluated in typical conditions (TT/\SI{0.80}{\volt}\kern-.16em/\SI{25}{\celsius}), with switching activities from a post-layout gate-level simulation running at \SI{600}{\mega\hertz}.

The area overhead of different implementations is shown in \cref{table:area}. Even optimized implementations of \lrscwait{} quickly grow in size, while \lrscwait{ideal} is physically infeasible for a system of \mempool{}'s scale. \colibri{}, on the other hand, grows linearly and allows up to eight queues per memory controller with a similar area overhead to \lrscwait{1} of \SI{16}{\percent}.

\begin{table}[t]
  \centering
  \caption{Area of a \texttt{mempool\_tile} with different \lrscwait{} designs.}
  \label{table:area}
  \vspace{-0.1cm}
  \begin{tabularx}{\columnwidth}{@{}Xlrr@{}}
    \toprule
    \textbf{Architecture}   & \textbf{Parameters} & \textbf{Area{[}kGE{]}} & \textbf{Area{[}\%{]}}  \\
    \midrule
    \mempool{} tile         & none          & 691       & 100.0   \\
    with \lrscwait{1}       & 1 queue slot  & 790       & 116.4   \\
    with \lrscwait{8}       & 8 queue slots & 865       & 127.4   \\
    with Colibri with MWait & 1 address     & 732       & 105.9   \\
    with Colibri with MWait & 2 addresses   & 750       & 108.5   \\
    with Colibri with MWait & 4 addresses   & 761       & 110.1   \\
    with Colibri with MWait & 8 addresses   & 802       & 116.3   \\
    \bottomrule
  \end{tabularx}
\end{table}

\subsection{Benchmarking}

\paragraph{Histogram}
We implement a concurrent histogram benchmark to evaluate \colibri{}'s performance at different levels of contentions. The application atomically increments a parametrizable number of bins. The fewer bins, the higher the contention. We increment a bin with different atomic operations and compare their performance as updates per clock cycle.

\begin{figure}[t]
    \centering
    \includegraphics[]{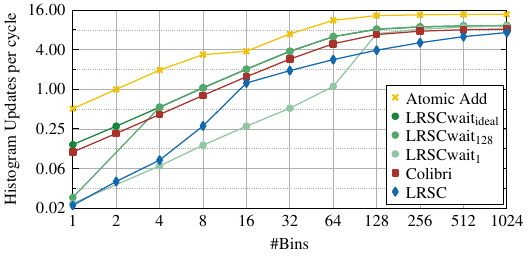}
    \caption{Throughput of different \lrscwait{} implementations and standard \riscv{} atomics at varying contention.}%
    \label{fig:results:histogram_atomics}
\end{figure}

The throughput of different \lrscwait{} implementations is shown in \cref{fig:results:histogram_atomics}. \lrscwait{ideal} outperforms all implementations across the whole spectrum of contention. The optimized implementations show similar performance at low contention but achieve much lower performance when the contention is higher than their number of reservations. Finally, \colibri{} achieves near-ideal performance across all contentions. The slight performance penalty comes from the extra roundtrips of \colibri{}'s node update messages. \colibri{} outperforms the \gls{lrsc}-based implementation by a factor of 6.5$\times$ at high contention and \SI{13}{\percent} at low contention. For completeness, we also show the throughput of an \emph{Atomic Add} implementation, which is designed specifically to increment a memory location atomically and represents the plot's roofline. However, most concurrent algorithms need more complex atomic \gls{rmw} operations than an increment, where programmers have to resort to locks of generic \gls{rmw} atomics like \lrscwait{}.

\begin{figure}[t]
    \centering
    \includegraphics[]{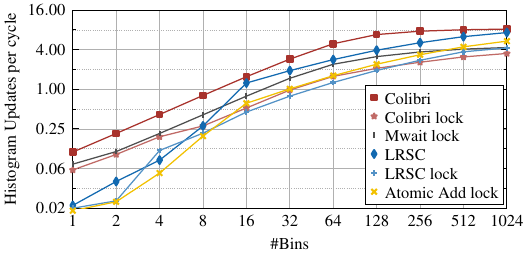}
    \caption{Throughput of different lock implementations compared to generic \gls{rmw} atomics at varying contention.}%
    \label{fig:results:histogram_locks}
\end{figure}

\cref{fig:results:histogram_locks} compares \colibri{} to various lock-based implementations. \colibri{}, \gls{lrsc}, and Atomic Add locks are spin locks with a backoff of 128 cycles, while \mwait{} lock implements an \gls{mcs} lock, where \mwait{} is used to avoid polling. \colibri{} outperforms all other approaches for any contention. We observe that the \lrsc{} and AMO-lock approaches perform worst at high contention due to their heavy polling and retry traffic, while waiting-based approaches perform average. At low contention, the waiting-based approaches perform worst because of their management overhead, while the other atomics tend to \colibri{}.

\paragraph{Interference}

\begin{figure}[t]
    \centering
    \includegraphics[]{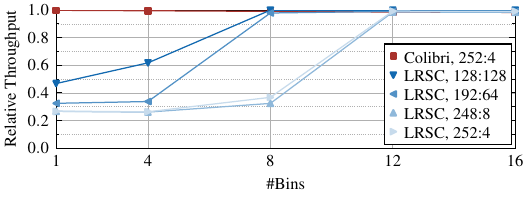}
    \caption{Matrix multiplication performance with interference from atomics. The poller-to-worker ratio is annotated in the figure with poller:worker.}%
    \label{fig:results:interference}
\end{figure}

We showed that \lrscwait{} can significantly improve the throughput of atomic operations across all levels of contention. On top of this increase in performance, eliminating the need to retry failed operations and polling also reduces traffic and frees up resources for cores not executing atomics. Cores working on computation experience less negative interference from the constant polling of atomics. To measure this effect, we partitioned the 256 cores of \mempool{} to either work on a matrix multiplication or to execute atomic operations. We measure the execution time of the matrix multiplication compared to an execution time without any interference. \Cref{fig:results:interference} shows the relative performance for various types of atomic operations and distributions of working cores. Our \colibri{} implementation has a negligible impact on the worker cores, even at high contention and with a poller-to-worker ratio of 252:4. The retries of the \gls{lrsc} operations, on the other hand, significantly impact the workers' performance, despite a backoff of 128 cycles. At the same ratio of poller-to-workers, the \gls{lrsc} implementation slows the workers down to \SI{26}{\percent}.

\begin{figure}[t]
    \centering
    \includegraphics[]{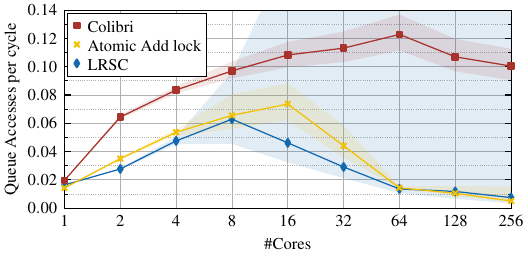}
    \caption{Queue operations throughput with different atomics.}%
    \label{fig:results:queue}
\end{figure}

\paragraph{Queue}
To evaluate \colibri{} on a commonly used concurrent algorithm, we implement an \gls{mcs} queue with \gls{lrsc} and \lrscwait{}, as well as a lock-based queue using atomic adds. Concurrent queues are widely used for task scheduling or producer/consumer pipelines. \Cref{fig:results:queue} shows the number of queue operations for a range of cores accessing a single queue. \colibri{} performs best and can sustain a high performance even at 256 cores. It outperforms the \gls{lrsc} and lock-based approaches by 1.54$\times$ and 1.48$\times$ times with eight cores before both implementations drop in performance due to excessive retries and polling. At 64 cores, \colibri{} is 9$\times$ faster. The shaded areas show each implementation's slowest and fastest core performance range. It illustrates how \colibri{} results in a very balanced and fair workload distribution, while \gls{lrsc} can have very big variations.

\paragraph{Energy efficiency}

\Cref{table:power_results} shows the energy per operation for atomic accesses to the histogram at the highest contention. Comparing \colibri{} to the Atomic Add, which represents an ideal atomic update, we can see how energy-efficient \colibri{} is for a generic \gls{rmw} operation that consists of an \lrwait{}, add, and \scwait{} operation. Compared to the \gls{lrsc} or lock-based implementation, we observe the large benefit of the reduction in polling and retry traffic for improving energy efficiency by a factor of 7.1$\times$ and 8.8$\times$.

\begin{table}[t]
  \centering
  \caption{Area results for a \texttt{mempool\_tile} for ideal LRWait.}
  \label{table:power_results}
  \vspace{-0.1cm}
  \begin{tabularx}{\columnwidth}{@{}Xrrrr@{}}
    \toprule
    \textbf{Atomic access} & \textbf{Backoff} & \textbf{Power (\si{\milli\watt})}  & \textbf{Energy (\si{\pico\joule\per\op})} & \textbf{$\Delta$} \\
    \midrule
    Atomic Add             & 0    & 175       &   29 &   $-$\SI{77}{\percent}  \\
    Colibri                & 0    & 169       &  124 & $\pm$\SI{00}{\percent}  \\
    \gls{lrsc}             & 128  & 186       &  884 &  $+$\SI{613}{\percent}  \\
    Atomic Add lock        & 128  & 188       & 1092 &  $+$\SI{780}{\percent}  \\
    \bottomrule
  \end{tabularx}
\end{table}

\section{Conclusion}
\label{sec:conclusion}

In this work, we propose the \lrwait{} and \mwait{} synchronization primitives and their implementation, \colibri{}, which demonstrate a novel and effective solution for the \gls{lrsc} synchronization problem in cache-less manycore systems. \colibri{} offers superior performance and scalability compared to existing hardware and software approaches, reduces polling, and improves throughput in a fair manner. Our experiments show that \colibri{} outperforms other implementations in both high and low contention scenarios by up to 6.5$\times$ and improved energy efficiency by up to 8.8$\times$.
The polling and retries of \gls{lrsc}-based solutions can lead to performance degradation of unrelated workers by up to 4$\times$, while \colibri{} can operate even at high contention without impacting other cores. 
Additionally, \colibri{} can be easily integrated into existing RISC-V systems with a small hardware overhead and can be used as a drop-in replacement for \gls{lrsc} or as an extension.

\section*{Acknowledgment}
This work is funded in part by the COREnext project supported by the EU Horizon Europe research and innovation programme under grant agreement No. 101092598.

\Urlmuskip=0mu plus 1mu\relax
\def\UrlBreaks{\do\/\do-}
\bibliographystyle{IEEEtran}
\bibliography{bib/lrwait}

\end{document}